\documentclass{mod_ws-mpla}

\newcommand{\bea}{\begin{eqnarray}}
\newcommand{\eea}{\end{eqnarray}}
\newcommand{\be}{\begin{equation}}
\newcommand{\ee}{\end{equation}}
\newcommand{\ba}{\begin{array}}
\newcommand{\ea}{\end{array}}

\newcommand{\lnum}{{\cal L}}
\newcommand{\bnum}{{\cal B}}
\newcommand{\md}{{\rm{~mod~}}}
\newcommand{\lmat}[2][lllll]{\left( \begin{array}{#1} #2\\ \end{array} \right)}

\newcommand{\rmat}[2][rrrrr]{\left( \begin{array}{#1} #2\\ \end{array} \right)}
\newcommand{\larray}[2][lllll]{\begin{array}{#1} #2\\ \end{array}}

\newcommand{\lsim}{
\mathrel{\hbox{\rlap{\hbox{\lower4pt\hbox{$\sim$}}}\hbox{$<$}}}}
\newcommand{\gsim}{
\mathrel{\hbox{\rlap{\hbox{\lower4pt\hbox{$\sim$}}}\hbox{$>$}}}}
\newcommand{\gev}{~\text{GeV}}
\newcommand{\ev}{~\text{eV}}
\newcommand{\mpl}{M_{Pl}}

\newcommand{\til}{\widetilde}
\newcommand{\phix}{\phi_X}
\newcommand{\psix}{\psi_X}

\begin{document}

\markboth{Hye-Sung Lee}
{$R$-parity violating $U(1)'$-extended supersymmetric standard model}

\catchline{}{}{}{}{}

\title{$R$-parity violating $U(1)'$-extended supersymmetric standard model}

\author{Hye-Sung Lee}

\address{Department of Physics and Astronomy, University of California\\
Riverside, CA 92512, USA\\
hyesung@ucr.edu
}

\maketitle



\begin{abstract}
Supersymmetry is one of the best motivated new physics scenarios.
To build a realistic supersymmetric standard model, however, a companion symmetry is necessary to address various issues.
While $R$-parity is a popular candidate that can address the proton and dark matter issues simultaneously, it is not the only option for such a property.
We review how a TeV scale $U(1)'$ gauge symmetry can replace the $R$-parity.
Discrete symmetries of the $U(1)'$ can make the model still viable and attractive with distinguishable phenomenology.
For instance, with a residual discrete symmetry of the $U(1)'$, $Z_6 = B_3 \times U_2$, the proton can be protected by the baryon triality ($B_3$) and a hidden sector dark matter candidate can be protected by the $U$-parity ($U_2$).
\end{abstract}

\ccode{PACS Nos.: 11.30.Er, 12.60.Jv, 95.35.+d.}

\section{Introduction}
\label{sec:introduction}
Though standard model (SM) has been very successful in describing the nature, there are some issues in the model, which leads to a consensus that it is not an ultimate theory to describe the particle physics.
Among them is the gauge hierarchy problem, which is a fine-tuning issue in radiative correction of the Higgs mass.
This can be most naturally addressed by supersymmetry (SUSY).

The following is a superpotential of the general supersymmetric SM before any extra symmetry is introduced.
\be
\begin{split}
W &= \mu H_u H_d + y_E H_d L E^c + y_D H_d Q D^c + y_U H_u Q U^c \\
&+ \lambda LLE^c + \lambda' LQD^c + \mu' LH_u + \lambda'' U^cD^cD^c \\
&+ \frac{\eta_1}{M} QQQL + \frac{\eta_2}{M} U^cU^cD^cE^c + \cdots .
\end{split}
\ee
We can see that lepton number ($\lnum$) violation and/or baryon number ($\bnum$) violation at renormalizable and nonrenormalizable levels (in second and third lines) are one of the most general predictions of SUSY.
This superpotential has some problems.
First, there is the so-called $\mu$-problem.\cite{Kim:1983dt}
The $\mu$ parameter (in first line) of the superpotential should be of electroweak (EW) scale to avoid any fine-tuning in the electroweak symmetry breaking.
Supersymmetry itself does not explain why the SUSY conserving sector parameter $\mu$ should have the same scale as the SUSY breaking sector, not a higher scale such as Planck scale ($\mpl$).

\begin{figure}[tb]
\centerline{
\psfig{file=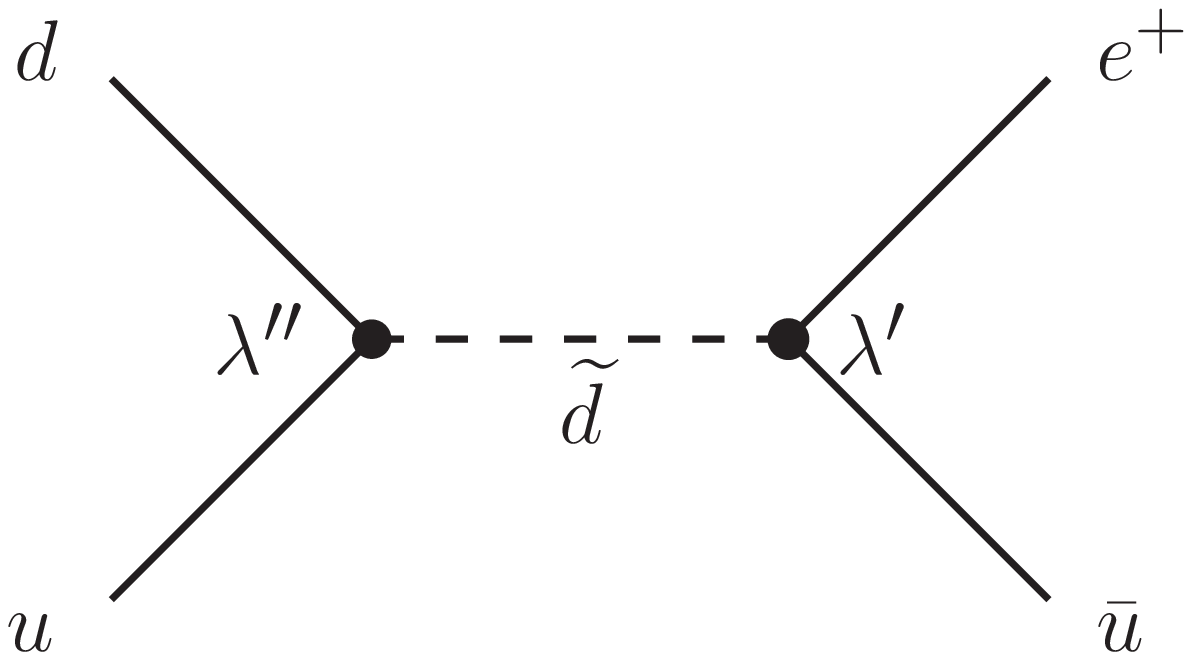,width=0.35\textwidth}
\psfig{file=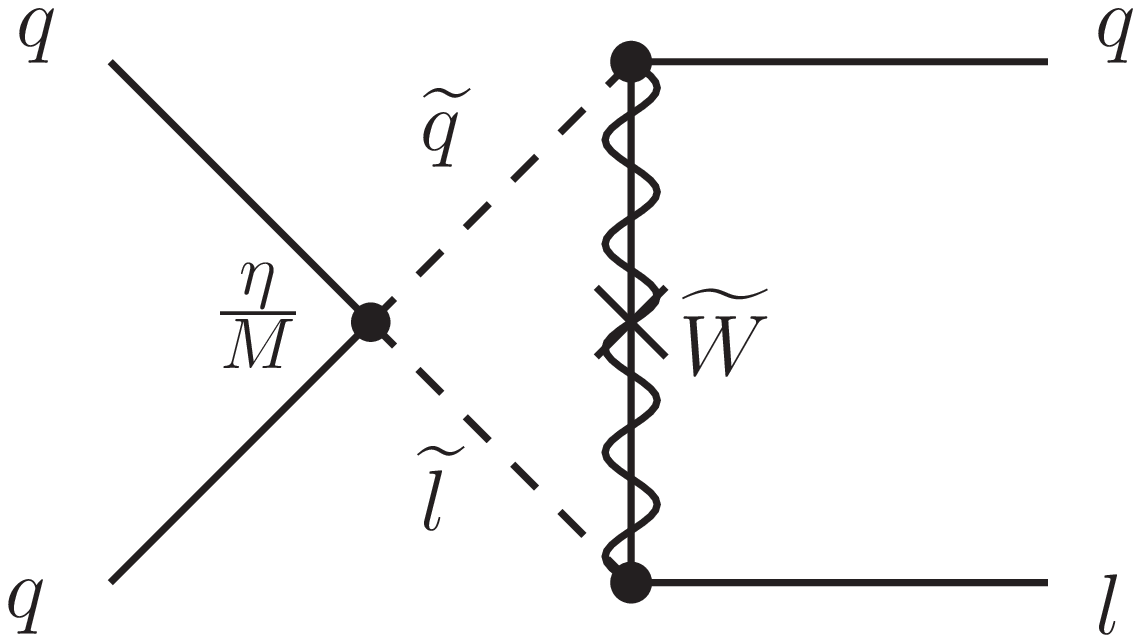,width=0.35\textwidth}
\psfig{file=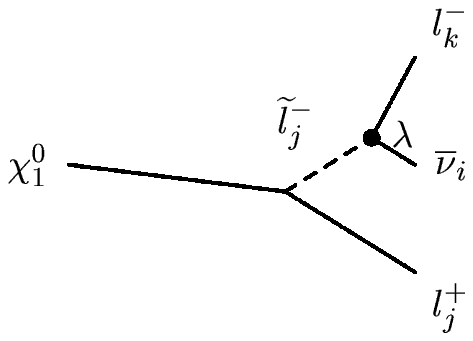,width=0.28\textwidth}}
\centerline{(a) ~~~~~~~~~~~~~~~~~~~~~~~~~~~~~~ (b) ~~~~~~~~~~~~~~~~~~~~~~~~~~~~~~ (c)}
\caption{(a) and (b) are proton decay diagrams through renormalizable and nonrenormalizable operators, and (c) is a neutralino LSP decay diagram.
\protect\label{fig:decay}}
\end{figure}

The $\lnum$ violating, $\bnum$ violating terms also cause problems with proton decay.
Figures~\ref{fig:decay} (a) and (b) are some examples of proton decay diagrams with dimension 4 operators and dimension 5 operators, respectively.
Proton decay requires both $\lnum$ violation and $\bnum$ violation.
To construct a proton decay diagram with dimension 4 operators, we need both $\lnum$ violating term and $\bnum$ violating term (e.g. $\lambda' LQD^c$ and $\lambda'' U^cD^cD^c$).
With dimension 5 operators, we can violate both of them with a single operator (e.g. $\frac{\eta}{M} QQQL$).
To satisfy a very long proton lifetime constraint ($\tau_p \gsim 10^{29} \mbox{ years}$),
the coefficients of the $\lnum$, $\bnum$ violating terms should satisfy the following constraints.
\bea
\mbox{Dimension } 4 &:&~ |\lambda_{LV} \cdot \lambda_{BV}| \lsim 10^{-27} , \\
\mbox{Dimension } 5 &:&~ |\eta| \lsim 10^{-7} 	\quad (\mbox{for } M = \mpl) .
\eea
For dimension 4 operators, if either $\lnum$ violating or $\bnum$ violating operator is absent, the other could have a sizable coefficient.
Dimension 5 $\lnum$ and $\bnum$ violating operators should have unnaturally small coefficients, even if we take the cut off scale $M = \mpl$, in order to ensure long enough life time for proton.\cite{Weinberg:1981wj}

Figure~\ref{fig:decay} (c) is a decay diagram of the lightest superparticle (LSP) with a neutralino example.
The LSP is the most popular dark matter candidate in supersymmetric models.
To be a viable dark matter candidate, however, the LSP lifetime should be similar to or larger than the Universe age ($t_0 \sim 14 \times 10^9 \mbox{ years}$).
It gives severe constraints of
\be
|\lambda|, |\lambda'|, |\lambda''| \lsim 10^{-20}
\ee
for TeV scale superparticle masses.

Therefore, it is clear that SUSY needs a companion symmetry to address these issues.
$R$-parity\cite{Farrar:1978xj} is a popular companion symmetry, but there are some aspects of the $R$-parity that are not completely satisfactory.
In this review, we will consider an alternative companion symmetry, a TeV scale Abelian gauge symmetry and see how it can play the roles that were played by the $R$-parity.
It will provide a solid theoretical framework to study the $R$-parity violating phenomenology with distinguishable predictions.

The rest of the paper is as follows:
In Section~\ref{sec:rparity}, we consider a popular SUSY companion symmetry, $R$-parity.
In Section~\ref{sec:u1}, we consider a TeV scale gauge symmetry, $U(1)'$.
In Section~\ref{sec:discrete}, we review the general residual discrete symmetry of an extra Abelian gauge symmetry.
In Section~\ref{sec:proton}, we discuss the residual discrete symmetry of the $R$-parity violating scenario and how it can help with the proton stability.
In Section~\ref{sec:lup}, we discuss the extension of the discrete symmetry to the hidden sector and a dark matter candidate in the $R$-parity violating scenario.
In \ref{sec:neutrino}, we review the mechanism for the neutrino to acquire mass in the $R$-parity violating $U(1)'$ model, after the summary and outlook in Section~\ref{sec:summary}.

\section{$R$-parity}
\label{sec:rparity}
Under $R$-parity, the SM fields have even parity and their superpartners have odd parity.
\be
R_p[{\rm SM}] = {\rm even} , \quad R_p[{\rm superpartner}] = {\rm odd} .
\ee
The $R$-parity is equivalent to matter parity in its effect.
While the $R$-parity is defined on the component fields, the matter parity is defined on the superfields.
Under matter parity, only matter superfields (fermions, sfermions) have odd-parity and the others (Higgses, gauge bosons, and their superpartners) have even parity.
\bea
\mbox{$R$-parity}    &:&~ R_p = (-1)^{3(\bnum - \lnum)+2s} , \\
\mbox{Matter parity} &:&~ M_p = (-1)^{3(\bnum - \lnum)} .
\eea
With an assumption of $R$-party, the LSP is absolutely stable, and it would be a good dark matter candidate if other conditions for a viable dark matter candidate are satisfied.

The general superpotential with $R$-parity is given as follows:
\be
\begin{split}
W_{R_p} &= \mu H_u H_d + y_E H_d L E^c + y_D H_d Q D^c + y_U H_u Q U^c \\
&+ \frac{\eta_1}{M} QQQL + \frac{\eta_2}{M} U^cU^cD^cE^c + \cdots .
\end{split}
\ee
Since $R$-parity does not address the $\mu$-problem, it still needs a separate solution or symmetry.
$R$-parity removes all renormalizable $\lnum$ violating terms and $\bnum$ violating terms, which is not really necessary for proton stability, forbidding potentially interesting phenomenology.\cite{Barbier:2004ez}
Furthermore, the $R$-parity does not prevent the dimension 5 $\lnum$ and $\bnum$ violating operators which still mediate too fast proton decay.
Therefore, the $R$-parity by itself is incomplete in addressing the proton stability.
So we need to look for an additional or alternative explanation or symmetry.

\section{TeV scale $U(1)'$ gauge symmetry}
\label{sec:u1}
We will consider a TeV scale $U(1)'$ gauge symmetry in this section.
When an extra Abelian gauge symmetry is introduced in the supersymmetric model, its natural scale is TeV scale.
Sfermions get extra $D$-term contributions from the $U(1)'$, and in order to make sure the soft terms are of TeV scale without fine-tuning, the $U(1)'$ should not be broken at larger than TeV scale.
For a $U(1)'$ breaking mechanism and other details of this model, we refer to a general review\cite{Langacker:2008yv} and references therein.

The TeV scale $U(1)'$ can provide a natural solution to the $\mu$-problem with the charge assignment that can forbid the original $\mu$ term ($\mu H_uH_d$) and allow an effective $\mu$ term ($hSH_uH_u$).
\be
z[H_u] + z[H_d] \ne 0 , \qquad z[S] + z[H_u]+z[H_d] = 0 . \label{eq:musolution}
\ee
$S$ is a Higgs singlet that breaks the $U(1)'$ spontaneously.\footnote{See Ref.~\refcite{Barger:2006dh} for a general discussion on the Higgs boson spectrum with an additional Higgs singlet.}
After $S$ gets a vacuum expectation value (vev) at TeV scale, the effective $\mu$ parameter is dynamically generated at EW or TeV scale.
\be
\mu_{\rm eff} = h\left<S\right> \sim {\cal O}(\mbox{EW/TeV}) .
\ee
This solution to the $\mu$-problem greatly motivates the TeV scale $U(1)'$ symmetry.

The mass of the new gauge boson $Z'$ is given by
\be
M_{Z'}^2 = 2 g_{Z'}^2 \left( z[H_u]^2 \left<H_u\right>^2 + z[H_d]^2 \left<H_d\right>^2 + z[S]^2 \left<S\right>^2 \right) ,
\ee
where $z[H_u]$, $z[H_d]$, $z[S]$ ($ \left<H_u\right>$, $ \left<H_d\right>$, $ \left<S\right>$) are the $U(1)'$ charges (vevs) of the Higgs fields $H_u$, $H_d$, and $S$, respectively.
The lower bound on $Z'$ mass by the Tevatron dilepton search is $M_{Z'} \gsim 700 \sim 1000 \gev$ depending on couplings.\cite{CDFdielectron}
Currently, the most stringent bound on $M_{Z'}$ is given indirectly by the primordial nucleosynthesis data,\cite{Barger:2003zh} and the LHC will be the first collider experiment that can probe this range (multi TeV) directly.

Consider the MSSM Yukawa terms,\footnote{We also include the right-handed neutrino. See \ref{sec:neutrino} for details.} (effective) $\mu$-term, and $[SU(2)_L]^2-U(1)'$ anomaly condition.
\be
\begin{split}
H_uQU^c &:~ z[H_u] + z[Q] + z[U^c] = 0 , \\
H_dQD^c &:~ z[H_d] + z[Q] + z[D^c] = 0 , \\
H_dLE^c &:~ z[H_d] + z[L] + z[E^c] = 0 , \\
H_uLN^c &:~ z[H_u] + z[L] + z[N^c] = 0 , \\
(S/M)^p SH_uH_d  &:~ (1+p) z[S] + z[H_u] + z[H_d] = 0 , \\
A_{221'} &:~ N_f (3z[Q] + z[L]) + N_H (z[H_u] + z[H_d]) + \delta = 0 .
\end{split}
\label{eq:conditions}
\ee
We used $\left(\frac{S}{M}\right)^p SH_uH_d$ to consider both the original $\mu$ term case ($p=-1$), and the effective $\mu$ term case ($p=0$).
$N_f$, $N_H$, $\delta$ mean the number of fermion families, the number of Higgs doublet pairs, and the exotic $SU(2)_L$ charged particle contribution to the $[SU(2)_L]^2-U(1)'$  anomaly.
In this review, we consider only the minimal fields assumption\footnote{See Ref.~\refcite{Hur:2008sy} for more general cases.} of
\be
N_f = 3 , \quad N_H = 1 , \quad \delta = 0 . \label{eq:minimal}
\ee

In order to have an anomaly free gauge theory, we should make sure the other anomaly conditions are also satisfied for a given particle spectrum. 
In our rather general treatment independent of specific spectrum, we do not consider these full gauge anomaly conditions in this review.
However, we refer to Refs.~\refcite{Cvetic:1997ky,Cheng:1998nb,Ma:2002tc,King:2005my,Lee:2007fw,Ma:2008wq} for some examples of anomaly free $U(1)'$ charge assignments.

We consider only the family universal $U(1)'$ charges for the MSSM sector.
There are 9 unknown $U(1)'$ charges ($Q, U^c, D^c, L, E^c, N^c, H_u, H_d, S$) and 6 conditions, which results in 3 free parameters in determining $U(1)'$ charges for the MSSM sector.
The general solution for the MSSM sector is then given by
\be
\lmat{
z[Q]\\
z[U^c]\\
z[D^c]\\
z[L]\\
z[E^c]\\
z[N^c]\\
z[H_u]\\
z[H_d]\\
z[S]}
=
\frac{\alpha}{3} \rmat{
 1\\
-1\\
-1\\
-3\\
 3\\
 3\\
 0\\
 0\\
 0}
+
\frac{\beta}{6} \rmat{
 1\\
-4\\
 2\\
-3\\
 6\\
 0\\
 3\\
-3\\
 0}
 +\frac{\gamma}{9} \rmat{
 (1+p)\\
8 (1+p)\\
- (1+p)\\
 0\\
 0\\
9 (1+p)\\
-9 (1+p)\\
 0\\
 9}
\label{eq:generalSol}
\ee
where the first vector (with $\alpha$) is the $\bnum - \lnum$, the second vector (with $\beta$) is the hypercharge ($y$).
Three free parameters can be written in terms of some $U(1)'$ charges as
\be
\alpha = z[H_d]-z[L] , \quad \beta = -2z[H_d] , \quad \gamma=z[S] . \label{eq:parameters}
\ee

Unlike the $\bnum$ violating term ($U^cD^cD^c$), there are more than one $\lnum$ violating terms at renormalizable level.
It is useful to note the conditions to have these terms are identical.
Since we already have
\be
y_E H_dLE^c , \quad y_D H_dQD^c , \quad h SH_uH_d ,
\ee
allowing the $\lnum$ violating terms
\be
\lambda LLE^c , \quad \lambda' LQD^c , \quad h' SH_uL
\ee
requires a common condition
\be
z[H_d] = z[L] .
\ee
In other words, if we want to control the $\lnum$ violating terms with a TeV scale $U(1)'$, we can only allow either all of them or none of them.

From Eqs.~\eqref{eq:conditions}, we can get the following relation:
\be
z[U^cD^cD^c] - z[LLE^c] + \frac{2}{3} z[H_uH_d] = 0 . \label{eq:LVBV}
\ee
The first term is a total $U(1)'$ charge of the $\bnum$ violating term ($U^cD^cD^c$), and the second term is a total $U(1)'$ charge of the $\lnum$ violating term ($LLE^c$).
The last term is proportional to a total $U(1)'$ charge of the original $\mu$ term ($H_uH_d$) or $-(1+p) z[S]$, which would be zero for $p=-1$ case, but nonzero for $p=0$ case (i.e. when the $U(1)'$ solves the $\mu$-problem).

When the $U(1)'$ solves the $\mu$-problem, according to Eq. \eqref{eq:LVBV}, if the $\bnum$ violating term exists (i.e. if the first term is zero), the $\lnum$ violating term cannot exist (i.e. the second term cannot be zero) since the last term is nonzero.
Similarly, if the $\lnum$ violating term exists, the $\bnum$ violating term cannot exist.
This means that the $\lnum$ violating terms and the $\bnum$ violating terms cannot coexist, which is called LV-BV separation.\cite{Lee:2007fw}
Therefore, the proton can not decay through the renormalizable level MSSM fields operators when the $\mu$-problem is solved by the $U(1)'$.

But there are more fields involved in this model.
The $[SU(3)_C]^2-U(1)'$ anomaly condition is
\be
3 (2 z[Q]+z[U^c]+z[D^c])+A_{331'}[\mbox{exotic colors}] = 0 .
\ee
The first term is equivalent to $-3 z[H_uH_d]$ due to the MSSM Yukawa relations, which would be required to be nonzero unless $p=-1$.
This means we need colored exotics that can contribute to this $[SU(3)_C]^2-U(1)'$ anomaly condition when the $\mu$-problem is solved by the $U(1)'$.\footnote{See Ref.~\refcite{Morrissey:2005uz} for a discussion about the gauge coupling unification in the presence of the $U(1)'$.}
We should address whether the proton can still be stable with the exotic fields and also at dimension 5 level.
We find that the residual discrete symmetry of the $U(1)'$ is a great tool for this argument, and we will utilize it in following sections.

It would be instructive to make a comment about the relation between the $U(1)'$ and $R$-parity.
In principle, it is possible to have the matter parity (equivalent to $R$-parity) as a residual discrete symmetry of the $U(1)'$ so that we can have both $R$-parity and the $\mu$-problem solution simultaneously from a single $U(1)'$ symmetry.\cite{Hur:2008sy}
In the $R$-parity conserving $U(1)'$ model, the LSP dark matter candidates can be new neutralino components\cite{Barger:2004bz,Barger:2005hb,Barger:2007nv} or even a sneutrino.\cite{Lee:2007mt}
In this review, however, we will consider only the $R$-parity violating case.

\section{Residual discrete symmetry of an Abelian gauge symmetry}
\label{sec:discrete}
$Z_N$ naturally emerges after $U(1)'$ is spontaneously broken by the vev of $S$.
After $U(1)'$ charges of all fields $z[F_i]$ are normalized to integers, the value $N$ and discrete charges $q[F_i]$ are determined by the following relations.
\bea
N &=& z[S] , \label{eq:NzS} \\
q[F_i] &=& z[F_i] \md N .
\eea
The Higgs singlet $S$ has $q[S] = 0$, which keeps the discrete symmetry unbroken after the $U(1)'$ symmetry is broken by the $\left<S\right>$.
As long as $N \ne 1$, there is generically a residual discrete symmetry of the $U(1)'$.

The most general $Z_N$ of the MSSM sector was first studied by Ibanez and Ross.\cite{Ibanez:1991pr}
The family universal $Z_N$ can be written as
\be
Z_N :~ g_N = B_N ^b L_N^\ell
\ee
with two generators
\be
B_N = e^{2\pi i \frac{q_B}{N}} , \quad L_N = e^{2\pi i \frac{q_L}{N}} .
\ee

There are 8 unknown discrete charges for $Q, U^c, D^c, L, E^c, N^c, H_u, H_d$.
We have 5 conditions from superpotential ($H_u Q U^c$, $H_d Q D^c$, $H_d L E^c$, $H_u L N^c$, $H_u H_d$), and another condition from the hypercharge shift invariance ($q[F_i] \to q[F_i] + \alpha y[F_i] \md N$).
Therefore, we have only 2 free parameters ($b, \ell$) to determine the MSSM sector discrete charges.

The family universal discrete charges of $B_N$ and $L_N$ as well as normalized hypercharges are given in Table~\ref{tab:discrete}.
As the table shows, the discrete charges are closely related to the $\bnum$, $\lnum$, and $y$, and the total discrete charge of $Z_N$ is given by
\be
q = b q_B + \ell q_L \md N = -(b \bnum + \ell \lnum) + b(y/3) \md N
\ee
with a conserved quantity $b \bnum + \ell \lnum \md N$.

We consider only the family universal discrete charges.
Family nonuniversal discrete charges are not likely at least in the quark sector since the mixing of quarks would not be allowed in contradiction to the expectation from the CKM matrix.
Family nonuniversal $U(1)'$ is still possible with the family universal $Z_N$ as long as the condition $z[F_i] = q[F_i] + n_i N$ is kept ($z[F_i]$ is family-dependent if $n_i$ is).\footnote{One should be careful since the family nonuniversal $U(1)'$ charges can cause a flavor changing neutral current by the $Z'$ in the physical eigenstate.\cite{Langacker:2000ju}
This flavor changing $Z'$, however, may explain the discrepancies in the rare $B$ decays.\cite{Barger:2003hg,Barger:2004hn} See also Refs.~\refcite{Barger:2004qc,Cheung:2006tm,Chiang:2006we} for its contribution to the $B$-$\bar B$ mixing.}


\begin{table}[t]
\tbl{Discrete charges of $B_N$ and $L_N$, and their relation with baryon number, lepton number, and hypercharge.}
{\begin{tabular}{@{}lccccccccl@{}} \toprule
 &  $Q$   & $U^c$ & $D^c$ & $L$   & $E^c$ & $N^c$ & $H_u$ & $H_d$      & meaning of $q$ \\
\colrule
$B_N$          & $0$   & $-1$  & $1$   & $-1$  & $2$   & $0$   & $1$   & $-1$       & $-\bnum + y/3$ \\ 
$L_N$          & $0$   & $~~0$   & $0$   & $-1$  & $1$   & $1$   & $0$   & $~~0$ & $-\lnum$ \\
$~y$            & $1$   & $-4$  & $2$   & $-3$  & $6$   & $0$   & $3$   & $-3$        & \\
\botrule
\end{tabular}
\label{tab:discrete}
}
\end{table}

\section{Discrete symmetry of the $U(1)'$ and the proton stability}
\label{sec:proton}
As we discussed in Section~\ref{sec:u1}, there are in general exotic fields in the $U(1)'$-extended supersymmetric standard model, which might change the discrete symmetry.
However, the MSSM discrete symmetry ($Z_N^{\rm MSSM} = B_N^b L_N^\ell$) still holds among the MSSM fields.
It is important to note that for a physics process which has only the MSSM fields in its effective operators (such as proton decay), we can still discuss it with the $Z_N^{\rm MSSM}$.

Now, we want to consider the $\lnum$ violation case.
We have another condition from the term $LLE^c$.
After the normalization to integers, we have the general $U(1)'$ charges for the $\lnum$ violating case as
\be
\lmat{
z[Q]\\
z[U^c]\\
z[D^c]\\
z[L]\\
z[E^c]\\
z[N^c]\\
z[H_u]\\
z[H_d]\\
z[S]}
=
I_Q \rmat{
 1\\
-4\\
 2\\
-3\\
 6\\
 0\\
 3\\
-3\\
 0}
+ 3 \rmat{
0\\
(1+p)\\
0\\
0\\
0\\
(1+p)\\
-(1+p)\\
0\\
1}
+
(1+p) \rmat{
 0\\
 1\\
-1\\
 1\\
-2\\
 0\\
-1\\
 1\\
 0}
\ee
with an integer $I_Q = z[Q]$.
The residual discrete symmetry should be a $Z_3$ symmetry since $N = z[S] = 3$ (see Eq.~\eqref{eq:NzS}).
Since the first column is just a hypercharge and the second column is an integer multiple of $N=3$, it is the last column that determines which $Z_3$ symmetry it is.
Its coefficient should be $(1+p) = 3 \cdot {\bf Z} \pm 1$ to have a discrete symmetry.
In the $p = -1$ case (where we have the original $\mu$ term, $H_u H_d$), it is not possible to satisfy this unless we change the minimal fields assumption of Eq.~\eqref{eq:minimal}.
In the $p = 0$ case (where we have an effective $\mu$ term, $SH_uH_d$), the condition is automatically satisfied.
Comparing the third column with general $Z_3 = B_3^b L_3^\ell$, we can see it is the $B_3$ symmetry, which is called baryon triality.\cite{Lee:2007qx}

From the discrete charge of $B_3$, $q = - \bnum + y/3 \md 3$, we have a selection rule
\be
B_3 :~ \Delta \bnum = 3 \times {\rm integer}
\ee
which prevents the proton decay ($\Delta \bnum = 1$) completely.\cite{Castano:1994ec}
In other words, any observation of the proton decay would invalidate the $B_3$ scenario.

For the $\bnum$ violating case, we have a condition from the $U^cD^cD^c$ term.
Following a similar argument, we will have the $L_3$ symmetry, which is called lepton triality.
The $L_3$ has a selection rule of 
\be
L_3 :~ \Delta \lnum = 3 \times {\rm integer}
\ee
which does not prevent the proton decay if the decay products has $3, 6, \cdots$ leptons.
However, it was shown that, with a particle spectrum specified, it can ensure proton stability up to dimension 5 level with a little help from the $U(1)'$ gauge symmetry.\cite{Lee:2007qx}
Any observation of the violation of the selection rule, such as neutrinoless double beta decay ($\Delta \lnum = 2$), would rule out the $L_3$ scenario.

\section{Discrete symmetry extended to hidden sector and the dark matter candidate}
\label{sec:lup}
The LSP is not a good dark matter candidate in general without the $R$-parity.
Though it is always possible to include an additional symmetry for a new dark matter candidate, e.g. $U(1)_{PQ}$ for axion dark mater, we will see if we can come up with a new dark matter candidate without introducing an independent symmetry.

When we try to satisfy the anomaly conditions with a new gauge symmetry, it is often necessary to include additional SM singlet fields.
They contribute to the $[{\rm gravity}]^2 - U(1)'$ and $[U(1)']^3$ anomalies.

For simplicity, we consider only Majorana type SM singlets ($X$), which get a mass of $U(1)'$ breaking scale with
\be
W_{\rm hidden} = \frac{\xi}{2} SXX .
\ee
This hidden sector field can be a dark matter candidate if it is stable.
The question is how to ensure the stability of this hidden sector field.
We consider a $Z_2$ parity, which we name $U$-parity, under which the MSSM fields have even parity, while the hidden sector fields have odd parity.\cite{Hur:2007ur}
\be
U_p[{\rm MSSM}] = {\rm even} , \quad U_p[{\rm hidden}] = {\rm odd} .
\ee
Then the lightest $U$-parity particle (LUP) would be either a fermionic or a scalar component of the $X$ field, which is stable due to the $U$-parity.
Now, the important part is that we do not want to introduce this hidden sector parity as an ad hoc addition, but we rather want it as a residual discrete symmetry of the $U(1)'$.

We introduce a new generator $U_N$ for the hidden sector discrete symmetry.
\be
Z_N^{\rm hid} :~ U_N .
\ee
This can only be $U_2$ for the Majorana type case,\footnote{For the Dirac type case and more general discussions, see Ref.~\refcite{Hur:2008sy}.} under which $q[{\rm MSSM}] = 0$ and $q[X] = -1$.

We take
\be
Z_N^{\rm tot} = Z_{N_1}^{\rm obs} \times Z_{N_2}^{\rm hid}
\ee
as a generalized residual discrete symmetry of the $U(1)'$ gauge symmetry extended to the hidden sector.\cite{Lee:2008pc}
$Z_N$ is isomorphic to $Z_{N_1} \times Z_{N_2}$, if $N_1$ and $N_2$ are coprime and $N = N_1 N_2$.
\bea
Z_N^{\rm tot} :~ g_N^{\rm tot} &=& B_{N_1}^b L_{N_1}^\ell \times U_{N_2}^u \\
&=& B_N^{b N_2} L_N^{\ell N_2} U_N^{u N1} .
\eea

\begin{figure}[t]
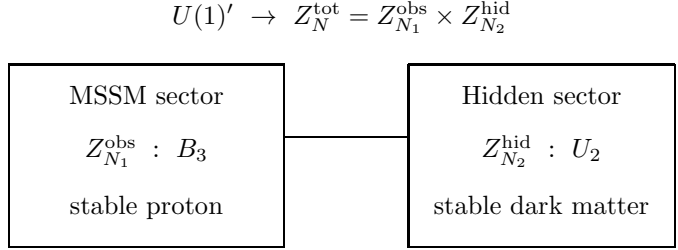

\begin{center}
\begin{tabular}{|c|c|c|}
\multicolumn{3}{c}{$U(1)' ~ \rightarrow ~ Z_N^{\rm{tot}} = Z_{N_1}^{\rm{obs}} \times
Z_{N_2}^{\rm{hid}}$} \\[4mm]
\cline{1-1}\cline{3-3} ~\hspace{30mm}~ &  ~\hspace{10mm}~ &
~\hspace{30mm}~\\[-2.5mm]
MSSM sector & & Hidden sector \\[3mm] 
\cline{2-2} && \\[-5mm]
$Z_{N_1}^{\rm{obs}}~:~ B_{3}$ && $Z_{N_2}^{\rm{hid}}~:~ U_{2}$ \\[3mm]
stable proton && stable dark matter \\[3mm]
\cline{1-1}\cline{3-3}
\end{tabular}
\caption{A single $U(1)'$ gauge symmetry provides the stability for the proton and the hidden sector dark matter candidate.}
\label{fig:picture}
\end{center}
\end{figure}

We consider the $\lnum$ violating case, as an example, where the MSSM sector has the $B_3$ symmetry as we discussed in the previous section.
With the $U_2$ in the hidden sector, the total residual discrete symmetry of the $U(1)'$ is $Z_6^{\rm tot} = B_3 \times U_2$.

As illustrated in Figure~\ref{fig:picture}, a unified picture about the stability of the proton and the hidden sector dark matter arises.
A $U(1)'$ gauge symmetry, which interacts with both the MSSM and hidden sectors, provides discrete symmetries to each sector.
In contrast to the usual $R$-parity scenario where a single discrete symmetry addresses the stability of the proton and the LSP dark matter candidate in the MSSM sector, in our case a single gauge symmetry, which we already have to solve the $\mu$-problem, addresses the stability of the proton in the MSSM sector and the LUP dark matter candidate in the hidden sector.

The total discrete charge of the $Z_6^{\rm tot}$ is given by $q = 2 q_B + 3 q_U \md 6$.
\be
\larray{
q[Q]=0 , & ~~~~q[U^c]=-2 , & ~~~~q[D^c]=2 , \\
q[L]=-2 , & ~~~~q[E^c]=-2 , & ~~~~q[N^c]=0 , \\
q[H_u]=2 , & ~~~~q[H_d]=-2 , & ~~~~q[X]=-3 .
}
\ee
Other possible exotic fields are assumed to be heavier than the proton and the LUP so that they are not stable due to the discrete symmetry.

It is useful to know that this $B_3 \times U_2$ in the $U(1)'$ gauge symmetry naturally arises without demanding it.
All one needs to do in order to have this in the minimal particle spectrum of Eq.~\eqref{eq:minimal} is to require 3 terms: (i) $SH_uH_d$ (i.e. solve the $\mu$-problem with the $U(1)'$), (ii) $LLE^c$ (i.e. demand a renormalizable $\lnum$ violating term), (iii) $SXX$ (i.e. demand an effective mass term for the Majorana type hidden sector field).
Then $B_3 \times U_2$ is automatically invoked as a residual discrete symmetry of the $U(1)'$ that ensures the stability of the proton and the LUP dark matter in this $R$-parity violating $U(1)'$ model (see Ref.~\refcite{Hur:2008sy} for details).

\begin{figure}[tb]
\centerline{
\psfig{file=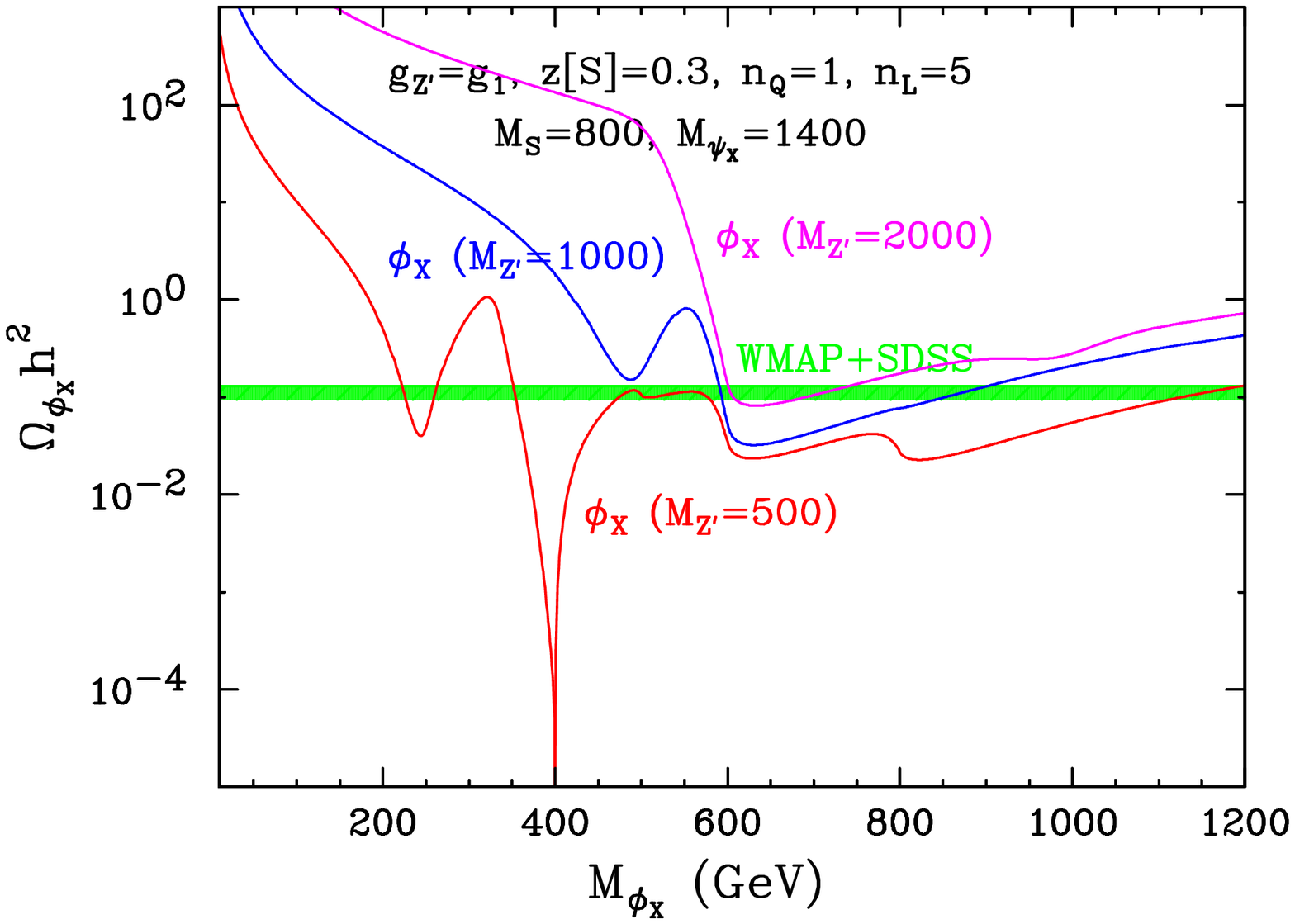,width=0.48\textwidth} ~~~
\psfig{file=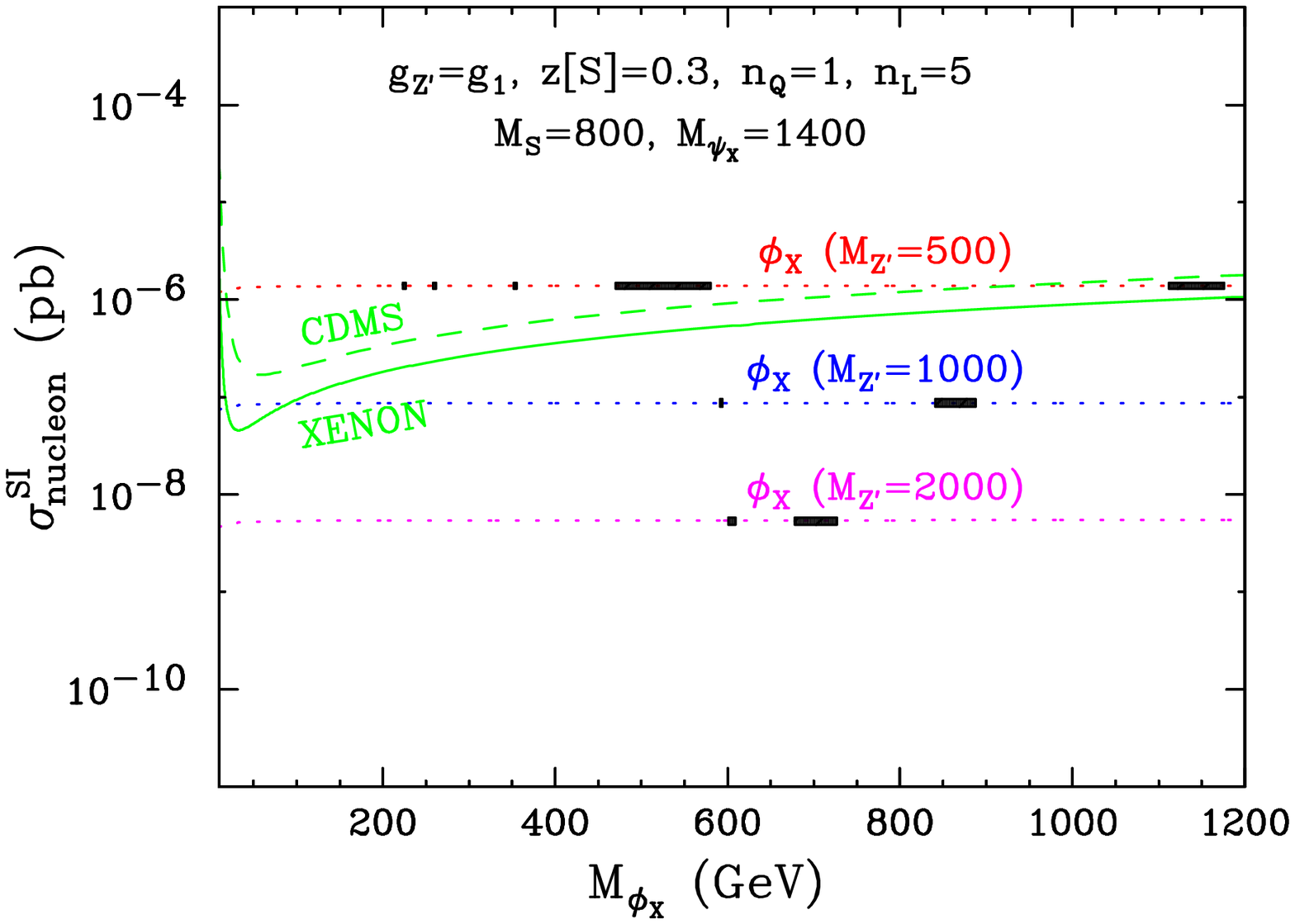,width=0.48\textwidth}}
\centerline{(a) ~~~~~~~~~~~~~~~~~~~~~~~~~~~~~~~~~~~~~~~~~~~~~~~~~~~ (b)}
\caption{(a) Relic density and (b) direct detection predictions for the scalar LUP ($\phix$) vs. $M_{\phix}$ with $M_{Z'} = 500$, $1000$, $2000 \gev$.
\protect\label{fig:LUPdarkmatter}}
\end{figure}

To be a viable dark matter candidate, however, the LUP should satisfy the constraints in the relic density and the direct detection.
The annihilation channels include, for the fermionic LUP ($\psix$) case,
\bea
\psix \psix &\to& f \bar f , ~S S , ~Z' Z' , ~S Z' , \\
\psix \psix &\to& \til f \til f^* , ~\til S \til S , ~\til Z' \til Z' , ~\til S \til Z' .
\eea
The annihilation channel to the superparticle pair, e.g. $\psix \psix \to \til f \til f^*$ ($S$ and $Z'$ mediated $s$-channel), is not possible in the usual LSP dark matter scenario unless the LSP and the next-to-LSP has tiny mass splitting.

Figure~\ref{fig:LUPdarkmatter} shows typical predictions of the relic density and the direct detection cross section of the LUP dark matter candidate.
For parameter values, see Ref.~\refcite{Hur:2007ur}.
They show the LUP dark matter can satisfy the experimental constraints from the WMAP+SDSS\cite{Spergel:2006hy} and the CDMS\cite{Akerib:2005kh}/XENON\cite{Angle:2007uj} simultaneously.
There can also be more than one discrete symmetries under which different dark matter candidates are stable, providing the multiple dark matters scenario naturally.
In this case, the cosmological constraints on the superparticles can be relaxed.\cite{Hur:2007ur,Cao:2007fy}

Therefore the LUP, which is a neutral, massive, and stable particle from the hidden sector, is a good and viable dark matter candidate.
It naturally arises when a new Abelian gauge symmetry is present in the model, and is a particularly attractive dark matter candidate in the $R$-parity violating model.

\section{Summary and Outlook}
\label{sec:summary}
While the supersymmetry is arguably best motivated new physics scenario, it requires a companion symmetry to be phenomenologically viable.
$R$-parity is most popular candidate for this companion symmetry, but it is not the only option.
Nevertheless, most of the SUSY phenomenology and its search schemes were developed based on the $R$-parity conservation.

We reviewed an alternative option, a TeV scale Abelian gauge symmetry $U(1)'$ with a certain residual discrete symmetry inside.
It is natural that a $U(1)'$ gauge symmetry has a residual discrete symmetry.
For example, $Z_6 = B_3 \times U_2$ can be invoked by just requiring the effective $\mu$ term, a lepton number violating term, and a Majorana type hidden sector mass term.
In this $R$-parity violating model, the $B_3$ (baryon triality) can ensure stability of the proton, and the $U_2$ ($U$-parity) can ensure stability of the hidden sector dark matter candidate (LUP).

This $R$-parity violating scenario with new TeV scale particle contents can open new possibilities in SUSY search schemes at the collider experiments as well as other distinguishable predictions.\cite{Lee:2008cn}

\appendix
\section{Neutrino mass}
\label{sec:neutrino}
We need to address the observed small neutrino mass ($m_\nu \lsim 0.1 \ev$).
In the $R$-parity violating $U(1)'$ model, there are several ways to have the neutrino mass.
\begin{romanlist}[(ii)]
\item Majorana neutrino with a usual seesaw mechanism\cite{Minkowski:1977sc,Yanagidaseesaw,Glashowseesaw,Mohapatra:1979ia,GellMann:1980vs}
\be
W = y_N H_uLN^c + m N^cN^c .
\ee
\item Dirac neutrino with a coupling that can be suppressed by the $U(1)'$\cite{Cleaver:1997nj,Langacker:1998ut}
\be
W = y_N \left(\frac{S}{M}\right)^a H_uLN^c .
\ee
\item Through lepton number violation without any right-handed neutrinos\cite{Hall:1983id,Grossman:1998py}
\be
W = \mu' H_uL + \lambda LLE^c + \lambda' LQD^c .
\ee
\end{romanlist}
While the $\lnum$ violating case can utilize any of three methods for neutrino mass, the $\bnum$ violating case can use only Dirac neutrino method since its discrete symmetry ($L_3$) does not allow any of $N^cN^c$, $H_uL$, $LLE^c$, and $LQD^c$.

\section*{Acknowledgments}
It is my great pleasure to thank all my collaborators.
This work is supported by the Department of Energy under grant DE-FG03-94ER40837.


\end{document}